# ARCHAIN: A Novel Blockchain Based Archival System

Albert Galiev, Shamil Ishmukhametov, Rustam Latypov,
Nikolai Prokopyev, Evgeni Stolov, Ilya Vlasov
Kazan Federal University, 35 Kremlevskaya st., Kazan, Russia
Email: crowbar_al@mail.ru, Shamil.Ishmukhametov@kpfu.ru, Roustam.Latypov@kpfu.ru,
nikolai.prokopyev@gmail.com, Yevgeni.Stolov@kpfu.ru, altavista@protonmail.com

*Abstract*—In this paper, we propose an novel archival system called ARCHAIN developed for the State archive-keeping committee of the Republic of Tatarstan (Russia). The blockchain is the primary part of the system, which stores transactions (facts of transfer of documents to the archive) in a protected form. The developed system uses a permission blockchain model due to the customer's requirement for the presence of a trusted center.

*Index Terms*—blockchain, archival science, cryptography

## I. INTRODUCTION

Blockchain is a novel kind of distributed ledger technology that uses cryptography to protect records during creation and storage. Initially, this technology was used in cryptocurrencies and although Bitcoin was the first of them, many alternatives have appeared with variations on the fundamental protocol and basic technology [1], [2]. Blockchain technology gives strength to cryptocurrency but it can be used for other purposes as well. Many authors predict widespread use of this technol- ogy beyond cryptocurrencies including financial transactions, recordkeeping etc [3], [4].

Recently, research and development have been done on application of blockchain technology in archives. A primary requirement for an archive is to set up mechanisms that will ensure such properties of it's contents as authenticity, integrity, trust in records in the long term. These properties can be provided to the system by using cryptographic primitives, including blockchain technology. In the paper [5] author presents a synthesis of original research conducted under the "Records in the Chain" project. In this paper the typology of blockchain solutions records managing as well as a few new archival projects aimed to leverage this innovative digital preservation technology has been presented. Authors of research [6] describe a novel approach to building of a decentralized, transparent, immutable and secure personal archive management system using blockchain technology.

This paper presents research and development performed by authors under the project "Use of blockchain technology in the field of implementation of acceptance of a scientific and technical documentation mandatory copy for archival storage" between October 2017 and December 2017. This project assumed the use of research results for experimental implementation of the transfer of a scientific and technical documentation mandatory copy to the archive using blockchain technology for the needs of the State archive-keeping committee of the Republic of Tatarstan (Russia). The paper describes concepts, principles and description of the software system, developed as a part of the project.

This paper is organized into four sections as follows:

1) Overview of blockchain technology
2) Technical task and problem statement
3) Possible attacks and security
4) System description

## II. OVERVIEW OF BLOCKCHAIN TECHNOLOGY

Blockchain is a distributed database technology for transaction processing. All transactions of blockchain are stored in a single register. Since all transactions are ordered by time the current state of the system is defined exclusively by this register. The main difference of blockchain from usual horizontally scalable databases such as MySQL Cluster is built-in security. It means computational impossibility to delete or change records in the database. Moreover, security is strengthened by the protocol of the blockchain itself.

Blockchains can be divided into types according to data access [7].

- Public blockchain is a blockchain in which there are no restrictions on data reading or sending of transactions in the blockchain.
- Private blockchain is a blockchain in which direct access to data and transactions sending is limited to a certain narrow range of participants.
- Permissionless (inclusive) blockchain is a blockchain in which there are no restrictions on the identity of transaction handlers (users who can create transaction blocks).
- Exclusive (permissioned) blockchain is a blockchain in which the processing of transactions is carried out by a certain list of identified persons.

Public blockchain technology (such as, for example, Bitcoin blockchain) hasn't been widely spread in data registration systems and in document assembly. The main reason for it is inability to control the processing of transactions (identities of transaction handlers should be known). Refusal from fundamental for public blockchains consensus reaching protocols allows to reduce the cost of system functioning. On the other side, many authors believe that exclusive blockchains are perspective in a way of being used by organizations for document automation. An exclusive blockchain can be either public or private.

Blockchain technology works with hashing algorithms and digital signatures. A digital signature is a primitive of asymmetric cryptography designed to provide authenticity of a digital message or a document. A digital signature gives grounds to the user to believe that:
- the message was created by an identified user,
- the sender can't refuse the fact of sending the message (non-repudiation),
- the message wasn't changed during the transmission (integrity).

## III. PROBLEM STATEMENT

The task of the project was to investigate the possibility of usage of blockchain technology in archival processes and to create a prototype of software for transfer of a scientific and technical documentation mandatory copy in an archive.

There were some customer restrictions on this system, for example, necessity of presence of a trusted center that distributes roles in the system network, as well as Russian Federation legislation requiring that Russian cryptographic standards must be used in cryptographic algorithms.

### A. Blockchain structure

Let us begin with definitions.

Administrator is an archive or an archive representative.

Signature is an electronic digital signature (EDS), formed by the EDS creation algorithm with use of a private key.

Administrator key (AK) is an open or private EDS key used by the Administrator to sign blocks at the moment of their creation and when making changes to the public key blockchain register.

Participants are organizations or organization representatives.

Participant key (PK) is an open or private EDS key used by the Participant when it is assigned by the Expert when the transaction is approved.

Expert is a person appointed by the Administrator to check the incoming transaction.

Timestamp is a sequence of characters that indicates an occurence of some event. It usually shows date and time.

Transaction is data unit containing transaction details and timestamp.

Public key certificates are used in the process of validating (verifying) digitally signed data when the recipient verifies that:
- the information identifying the sender was consistent with data contained in the certificate;
- no certificate from the chain of certificates was revoked and all certificates were valid at the time of signing the message;
- the certificate was used by the sender for its intended purpose;
- data hasn't been changed since the EDS was created.

The recipient can receive data signed by the sender after verification.

There are several subprograms with specific purpose:
- The program for a timestamp forming is a code fragment that adds a timestamp into the block.
- The program for forming a block of blockchain is a code fragment that uses a hashing program to form a new block.
- Verifier 2 is a certificate expiration date program.
- Verifier 3 is the program for checking of the main blockchain.
- Verifier 4 is the program to verify a result of a block creation in the blockchain.
- The hash program is a program for computing the hash of data.
- The signing program is a program for generating digital signatures using a private key.
- The program for the formation of a private and public key.
- The previous hash is the value of the hash function (hash) which was written in the previous block.
- The block hash is the value of the hash function obtained from the timestamp, transaction details and expert EDS of the previous hash.

Certificate Authority (CA) is a trusted entity that issues Digital Certificates and public-private key pairs. The Certificate Authority performs following functions:
- sets roles of Administrator, Participant, Expert;
- distributes private and public keys for authentication;
- validates the identity of the entity who requested a digital certificate;
- maintains THe Certificate Revocation List (CRL). It is a list of digital certificates which are no longer valid and therefore should not be relied on by anyone.

The blockchain is implemented as a table with three columns (Table I), where each line represents a separate transaction. The first column stores the transaction timestamp, the second column stores the details of the transaction with the digital signature of the authorized expert, and the third column stores the hash of the current transaction and its details with the hash of the previous transaction. After approval, the expert signs the transaction details and send them to the administrator, who then generatesthea timestamp and calculates the hash for the timestamp, signature details, and the hash from the previous transaction. When a new transaction is inserted into the block, the new block is distributed to participants. Since everyone knows the last hash, everyone can check that data hasn't been changed.

TABLE I
THE SUGGESTED ARCHITECTURE OF BLOCKCHAIN IN ARCHIVES.

| Timestamp | Blockchain title + Administrator EDS | |
|---|---|---|
| Timestamp | Transaction details + Expert EDS | Hash |
| Timestamp | Transaction details + Expert EDS | Hash |
| . . . | . . . | . . . |

In our project we use algorithms of electronic digital signature and hashing from the family of cryptographic hash functions called "Stribog", project name for GOST R34.10-2012 and GOST R34.11-2012 [8], [9].

## IV. POSSIBLE ATTACKS AND SECURITY

While working under "Database As a Rule" principle, a password system is viewed as the main tool to database security and reliability. But there is a way to interfere normal functioning of a database – corruption of database files. This malicious operation can be revealed with help of systematic observation of the state of database files. One calculates hashes of files and compares them to specific templates that are stored in some reliable place. This approach is impossible if a new data can be inserted at any time. What's more, a new problem arises – how to keep the integrity of templates. Usage of blockchain files can partially solve that problem. On the other hand, simple structure of database provides an opportunity to hide some changes in data. The structure of database files for other models can be rather complicated while in case of blockchain database any record can be selected very easily. In this section we demonstrate our approach to the security problem appeared during designing of the archive database.

The archive blockchain database is intended for keeping records of small size. Those can be diplomas, passports, and information about documents, placed into main storage, therefore all records in database contain information about transactions. In this way, placing a transaction into database means adding a record relating to that transaction in database. Those transaction does not contain any secret data, so the nature of any attack to the database can be:

1) Change of a record in the database;
2) Removing a record from the database;
3) Adding a record to the database bypassing the blockchain administrator.

There is a very important point related to the database integrity verification – the span of time between a malicious operation and the discovery of the attack. At this time intruder can use incorrect data extracted from the corrupted base and since the data obtained from official database has a proved status, some malicious actions on database can be undertaken. The fact of intrusion will be detected later, but the intruder will be out of access.

There are only two types of persons that have an opportunity to change content of the database: the administrator of the blockchain and administrators of servers where the database is deployed, so at least one of them must be involved into any attack. Of course, an illegal intrusion into the database from outside is also possible, but the problem of server protection from such attacks has no specialty in our case and it will not be studied in this paper. Now, let's consider each of attack types described above more thoroughly.

### A. Change of a record

According to requirements, the blockhain administrator can't change any transaction. If one has to change the doc- ument related to the transaction, the administrator creates a new transaction performing the procedure mentioned above. If the administrator ignores duties of the expert, which has to approve the document, this formed transaction will not be signed by any expert. To prevent this from happening, one additional feature of the program that inserts transaction into database must be implemented. Each expert that approves a document must have a second key that provides access to the procedure for placing the transaction. It means that such attack can be performed only under collusion between the administrator and the expert.

### B. Removing a record

Only the server administrator can do it. The standard pro- cedure for checking of database integrity will detect this event very easily. The only question is who will be notified about this situation. The database administrator or a supervisor of higher rank can be the person who starts the regular procedure for checking of integrity. The person who started the procedure must be aware about the result of the procedure. Although in that case a kind of collusion can have a place too, the time between the intrusion and it's detection has critical value.

### C. Adding a record while bypassing the regular procedure

From our point of view, this is the most real danger related to usage of blockchain database. At the time the last record is added to the base, this last record is not protected. The server administrator can replace this record with any other while just keeping the hash of the previous record. The next record will be added by the regular procedure, and the malicious operation will be detected only after analysis of content of the record. It takes a lot of time. To make such attack impossible we suggest to modify the regular procedure by adding some special operations. First of all, one develops a very fast procedure for calculation of hash of a big file. The hash is not very reliable but it is intended for keeping information between two sequential addings of records. While creating the first record of the blockchain, one calculates hash of that record and saves the hash in a file. The modified procedure for adding a new record to the database is as follows:

1) Create hash of the database, and compare the hash with the one in the old file.
2) If the calculated hash does not equal to the one in the old file, then alarm notification arises.
3) Extend the blockchain by the new record, delete the old file and create a new file with a new hash calculated on base of the extended blockchain.

The problem is who keeps this file with the hash. If blockchain administrator does it, then any malicious oper- ation can be performed only under collusion between both blockchain and server administrators. This is a more reliable solution but it is not very convenient since the file must be transferred twice during any transaction. The other solution uses an automatic procedure. Let us call old file as secret file. It means that the name of this file is generated on base of the content of the blockchain. The name looks like a random sequence, and the file is stored somewhere in a system folder of the main server. The only protection from the server administrator malicious actions in that case is opacity of the database manipulation for the server administrator.

The algorithm for creation of the secret file is presented be low (Algorithms 1, 2). Since blockchain file has byte structure, all operations are also byte oriented. The procedure $OneStep$ performs calculation $StateNew = Matr \cdot StateOld \oplus Byte$. All operations here are performed in $GF(2)$ field. $Matr$ is $8 \times 8$ binary matrix stored as byte array of length 8.

Two companion matrices of two primitive 8 degree polynomials are chosen in the main algorithm. Those matrices correspond to polynomials with coefficients 100011101 and 110101011 respectively [10].

## V. System description

The system was developed for the State archive-keeping committee of the Republic of Tatarstan to solve the problem described above. Functions of the archive include acceptance of documents for storing and information support on already accepted documents. To provide these features the system implements a public blockchain of transactions that contain information about the accepted documents. This blockchain will make it possible to verify the acceptance of the document at the specified time while also containing a part of the document details and information about network members who participated in transfering this document to the system.

### A. Blockchain implementation

Each accepted document corresponds to one transaction record in the blockchain. This record entry consists of:

1) Transaction timestamp
2) Document creation timestamp
3) Document metadata
4) Signature of the participant who created the document
5) Document examination timestamp
6) Signature of the participant who examined the document
7) Signature of the participant who added the document to the Archive
8) Transaction data imprint
9) Cryptographic link to the previous transaction (except for the first transaction)

The genesis block is a transaction which is signed by the administrator only. The genesis transaction doesn't contain nor data about the document neither a link to the previous transaction. Data imprints are implemented by the use of the cryptography standard "Stribog" [9]. Signatures are implemented by the use of the cryptography standard GOST 34.10- 2012 [8].

### B. Roles of participants

The system can be described as interaction of participants of three roles: Administrator, Expert, and User. Roles are selected and assigned to members through the Certification Authority.

Users create and upload documents to the network. Administrators select an expert for each of created documents and add them to the archive after expert's approval. Experts make decisions on documents – if the document is improperly formalized, has some metadata missing or doesn't comply to some local legislation, then it should be denied from transfering to the archive.

### C. Document management algorithm

The life cycle of the document in the system from the moment of creation to archival transfer can be described as a sequence of document status changes. A document can have one of the following statuses:

1) "Created" – this status is assigned after the user uploads the document to the network.
2) "On examination" – for each of documents the Administrator selects the Expert who must make a decision on the document.

---

**Algorithm 1** $OneStep(Matr, State, Byte)$

**Require:** $Matr$ - byte array of length 8; $State, Byte$ - bytes
 $Out \leftarrow 0$
 $Strbyte \leftarrow ToString(State)$ {Convert to string of length 8}
 **for** $i = 0$ to $7$ **do**
  **if** $Str[i] = 1$ **then**
   $Out \leftarrow Out \oplus Matr[i]$ {Modulo 2 operation}
  **end if**
 **end for**
 $Out \leftarrow Out \oplus Byte$
 **return** $Out$

---

**Algorithm 2** $Create\ secret\ file$

$SignLen \leftarrow NumbS$ {Any odd number} $Signature[SignLen]$ {Create byte array of length $SignLen$ padded by zeros}
$NameLen \leftarrow NumbM$ {Any number less than $SignLen$}
$FileName[NameLen]$ {Create byte array of length $NameLen$}
$Matr[2] \leftarrow Matr0, Matr1$ {Initialization of two byte arrays of length 8}
$State \leftarrow 0$
$Pos \leftarrow 0$
$MatrNumb \leftarrow 0$
**while** ($Byte = read(Blockchain) \neq EOF$ **do**
 $State \leftarrow OneStep(Matr[MatrNumb], State, Byte)$
 $MatrNumb \leftarrow (MatrNumb + 1)\%2$
 $Signature[Pos] \leftarrow State$
 $Pos \leftarrow Pos + 1)\%SignLen$
**end while**
$FileName\emptyset \leftarrow Signature[0:NameLen]$ {Convert part of signa- ture to secret file name}
write $Signature$ to file $FileName$

3) "Approved" – the Expert consider that the document was properly formed, contains all data required and it is ready to be transferred to the archive.
4) "Rejected" – the Expert consider that the document is not ready to be transfered to the Archive. The User must create a new document with all changes needed.
5) "Expired" – a fixed time is given for examination after which the status of the document will be changed automatically, and the Expert will lose the opportunity to make a decision on the document.
6) "Added" – after receiving the Expert's verdict, the Administrator transfers the document to the archive.

Each document status change is propagated through the system as an intermediate transaction signed by the appropriate role: the status "Created" is signed by the User, statuses "Approved" and "Rejected" are signed by the Expert, statuses "On examination", "Expired", and "Added" are signed by the Administrator. Intermediate transactions are available to all participants of the network, but their role in the system is formal as they are not added into the blockchain. After receiving the "Added" status, the final transaction on the document is formed and added to the blockchain.

### D. Software implementation

The User and Expert roles are implemented as desktop client applications.

The Administrator role is implemented as two applications: the desktop client and the server administrative node. Administrative nodes are used to store document data before transferring it for storage on archive servers and to connect users to each other. Network members can also connect directly to each other.

Desktop client applications are implemented using the Electron framework, the administrative node is written in Golang.
To connect network members, the HTML5 WebSockets protocol is used, as it provides an easy two-way exchange of arbitrary data. Due to the restrictions imposed on cryptography standards in the state institutions of the Russian Federation, an unprotected version of the websocket protocol is used, but security and encryption are provided at the application level using the standards of Russian cryptography.

For the system to work, it is necessary to provide a trusted channel of data exchange between the certification authority and the administrative node. In the implemented prototype, this channel was provided with the use of REST API on the side of the administrative node paired with authorization tokens.

### E. Certification Authority

Trust in the system is ensured by a centralized Certification Authority. Such a decision, of course, forces us to abandon a part of decentralized nature of the blockchain technology. However, in this way system participants are rid of the need for expensive calculations, meanwhile every member of the network retained the ability of verification performed under each document transaction. The certification authority distributes information on current valid certificates to the participants of the network. Thus, any node can verify all connected network members, and any network member can verify the authenticity of transactions.

The main tasks of certification authority are:
- Registration of new participants in the network;
- Edition of public key certificates for users and administrators of the system;
- Support for active and expired certificates;
- Checking for the authenticity and integrity of certificates;
- Notifying to system participants of the certification authority blockchain updates.

The Certification Authority is implemented as an ASP.NET application developed with C# programming language. PostgreSQL was used to store data.

### F. User registration

Uer passwords are, as usual, stored in the database in form of hashes. Here in the system we used the Russian standard GOST R 34.11-2012, called "Stribog" for hashing. After the authorization is completed, the user has an opportunity to enter personal profile settings and specify additional information about first name, last name, organization, e-mail etc.

Each member of the network plays its own role. Two new roles was defined in the Certification Authority in ad- dition to the ones mentioned above:an Unconfirmed User and Certification Authority Administrator (CA Administrator). An Unconfirmed User doesn't have access any of applications and can't' own the certificate. The main task of the CA Administrator is distribution of roles for all users registered in the system. He is provided with the list of all registered users with their role select controls.

### G. Certificate support

If the CA Administrator assigns user a category other than an Unconfirmed User category, the user's public key certificate will be automatically generated based on user profile data. It also happens when a not unconfirmed user edits his own profile data. A certificate is a text file that contains the following data:
1) Unique number of the certificate;
2) Unique number of the certificate holder;
3) Full name of the certificate holder;
4) The e-mail address of the certificate holder;
5) The certificate holder organization;
6) The certificate holder category;
7) The certificate holder public key in the form of a 1024-bit number in accordance to GOST R 34.10-2012;
8) Date of certificate expiration;
9) Name of the algorithm used for public key generation;
10) Metadata about the Certification Authority.

This text file can be downloaded by his owner at any time. The Russian standard GOST R 34.10-2012 is used for electronic digital signature creation. This algorithm is based on elliptic curves in a finite field. In our implementation we used an EDS algorithm with a key length of 512 bit. The public key is the point $P(x; y)$ of the predefined elliptic curve $E$, where $x, y$ are 512-bit coordinates of the point $P$.

TABLE II
THE ARCHITECTURE OF BLOCKCHAIN OF ALL CERTIFICATES.

| Timestamp | Certificate hash + CA Administrator EDS | Hash |
|---|---|---|
| Timestamp | Certificate hash + CA Administrator EDS | Hash |
| Timestamp | Certificate hash + CA Administrator EDS | Hash |
| ... | ... | ... |

The elliptic curve $E$ and the point $Q$ lying on it are chosen unanimously. The private key is the 512-bit number $k$, such that $P = k\,Q$. The Certification Authority Administrator has its own pair of keys, the validity of which is unlimited in time.

Public and private keys are generated in process of new certificate creation. The public key is stored in the certificate, and the private key is shown to the user only once, so it is very important not to lose it. The certificate data is written to a special blockchain of all certificates after the certificate is created. It is a blockchain in which the hash values of all previously created certificates are stored, as well as the signature of the CA Administrator (Table II).

After building of a new block in the blockchain of all certificates, the blockchain update is sent to the Administrator server node. If the user already had an active certificate, we create a new block for the old certificate in the other blockchain. It is the blockchain of revoked certificates built in a similar way to the blockchain of all certificates. Thus, checking of the validity of a certificate comes down to finding it's hash value in the blockchain of revoked certificates, and then in the blockchain of all certificates. If it is found in the blockchain of revoked certificates, the certificate will be decided as invalid. If it is not found there, but it is found in the blockchain of all issued certificates, the certificate will be decided as valid.

If the certificate expires, it will be automatically revoked and it's owner will be prompted to create a new one. If the CA Administrator changes user's category to the Unconfirmed User category, the certificate will be revoked without the right to receive a new one until user will have recieve a not Unconfirmed User category.

## VI. CONCLUSION

ARCHAIN is designed to conduct an experiment on the acceptance by archive of a scientific and technical documentation mandatory copy. New methods of protection against cheating at the time of a new block creation are proposed, while the participants consensus algorithm is not utilized on account of functioning under conditions of a permissioned blockchain usage. Further development involves the deployment of the software for use in the whole archival system of the Republic of Tatarstan, as well as it's integration into the system of record keeping.

## REFERENCES

[1] S. Nakamoto, "Bitcoin: A Peer-to-Peer Electronic Cash System", https://bitcoin.org/bitcoin.pdf
[2] A List Of Altcoins https://www.investitin.com/altcoin-list/
[3] S. Underwood, "Blockchain Beyond Bitcoin" Communications of the ACM, vol. 59 no. 11, pp. 15–17.
[4] Z. Zheng, S. Xie, H. Dai, X. Chen, and H. Wang An, "Overview of Blockchain Technology: Architecture, Consensus, and Future Trends", 2017 IEEE 6th International Congress on Big Data, Honolulu, HI, 2017, pp. 557–564.
[5] V.L. Lemieux, "A Typology of Blockchain Recordkeeping Solutions and Some Reflections on their Implications for the Future of Archival Preservation", 2017 IEEE International Conference on Big Data, Boston, MA, 2017, pp. 2271–2278.
[6] Z. Chen and Y. Zhu "Personal Archive Service System using Blockchain Technology: Case Study, Promising and Challenging", 2017 IEEE International Conference on AI & Mobile Services, Honolulu, HI, 2017, pp. 93–99.
[7] H. Okada, S. Yamasaki, and V. Bracamonte, "Proposed classification of blockchains based on authority and incentive dimensions", 19th International Conference on Advanced Communication Technology, PyeongChang, 2017, pp. 593–597.
[8] GOST R 34.10-2012: Digital Signature Algorithm https://tools.ietf.org/html/rfc7091
[9] GOST R 34.11-2012: Hash Function https://tools.ietf.org/html/rfc6986
[10] A. Gill, Linear sequential circuits, McGraw-Hill, 1966.